# Decomposing an information stream into the principal components

A.M. Hraivoronska, D.V. Lande*,

Institute for information recording of NAS of Ukraine

*dwlande@gmail.com

*We propose an approach to decomposing a thematic information stream into principal components. Each principal component is related to a narrow topic extracted from the information stream. The essence of the approach arises from analogy with the Fourier transform. We examine methods for analyzing the principal components and propose using multifractal analysis for identifying similar topics. The decomposition technique is applied to the information stream dedicated to Brexit. We provide a comparison between the principal components obtained by applying the decomposition to Brexit stream and the related topics extracted by Google Trends.*

## 1 Introduction

Analytical investigation of information from the Internet provides such pressing challenges as identifying trends in the information spread process as well as detecting anomalies in it. In such cases, thematic information streams are often studied [1]. The thematic information stream consists of documents dedicated to the given topic. Each document has a timestamp (most often a publication date). Such information streams are gathered using a content-monitoring system.

On the one hand, to reflect the dynamics of an information stream we use time series that consists of the numbers of publications during a specific time unit. Then, to study dynamics, we apply nonlinear analysis methods such as wavelet-, fractal, and multifractal analysis [2, 3].

On the other hand, documents in an information stream are often more narrowly focused than the general topic. Therefore, the documents can be quite heterogeneous. For example, choose a subject of scientific publications as the main topic, and then the keywords of the articles can serve as the narrow topics. Another example refers to news publications. In the information stream dedicated to Brexit, there are publications on economic, social, and other consequences of secession from EU, in particular about immigration and trade, as well as simple messages about the forthcoming referendum and the results of past referendum [4, 5].

One may consider the thematic information stream as a set of information streams dedicated to more narrow topics. Exploring the roles of subtopics in the collection of documents allows describing the structure of the information stream in more detail. Thus, modeling the information stream with due regard to subtopics in its documents is a current task.

Developing method for the decomposition is a topical problem because such method allows reducing the complicated problem of analytical processing of information streams. Such technique can be applied to solve following problems: identifying primary, minor, and background insignificant topics in the document collection; presenting additional information for recognizing crucial moments of information influence due to splitting the information process into components.

Note several approaches to the analysis of topics in a text corpus. In the field of natural language processing, there is the topic modeling subject. In the context of text modeling, each document is dedicated to a specific topic, and each topic is characterized by specific words or word collocations. The first probabilistic thematic model named pLSI (probabilistic Latent Semantic Indexing) was proposed in 1999 [6-8]. According to the pLSI model, each word in a text is a sample from a mixture model. The mixture components are multinomial random variables, which is the representation of topics. A few years later, the LDA (Latent Dirichlet Annotation) model was proposed. The LDA and pLSI models are quite similar. The difference is that the prior topic distribution is the Dirichlet distribution in the LDA.

The approach described in this article focuses on the dynamics of an information process, which is its key feature. The article is organized as follows. In Section 2 we introduce the decomposition of the information process into the principal components and show the analogy with the Fourier transform. In section 3 we propose methods for analyzing the principal components obtained through the decomposition. A practical example of applying the decomposition to the Brexit information stream is provided in Section 4. In Section 5 we compare the principal components for the Brexit topic and the related topics extracted by Google Trends.

## 2 A process of decomposing the information stream

### 2.1 Mathematical description of the decomposition

We propose to decompose the information stream in the similar way as the Fourier transform decomposes a signal. The essential idea is to extract to some extent simple components from a complex object with a time domain. In Fourier analysis, we decompose a function of time into the oscillatory components with different frequencies

$$f(t) = \frac{a_0}{2} + \sum_{k=1}^{\infty} (a_k \cos kt + b_k \sin kt).$$

The Fourier transform is an approach to get from the time domain to the frequency one and, as a result, obtain additional hidden information about the signal. In the case of the decomposition, analogs of frequencies are narrow topics.

The idea of the information stream decomposition is illustrated in Fig. 1. The dynamics of numbers of publications dedicated to the main topic is at the top of the figure. Looking at the dynamics, we have no idea about what kinds of messages making it up. By applying the decomposition, we aim to extract the hidden information and present it in a way useful for further analytical investigation, namely, time series corresponding to essential subtopics. These time series are schematically illustrated below the arrow in Fig. 1.

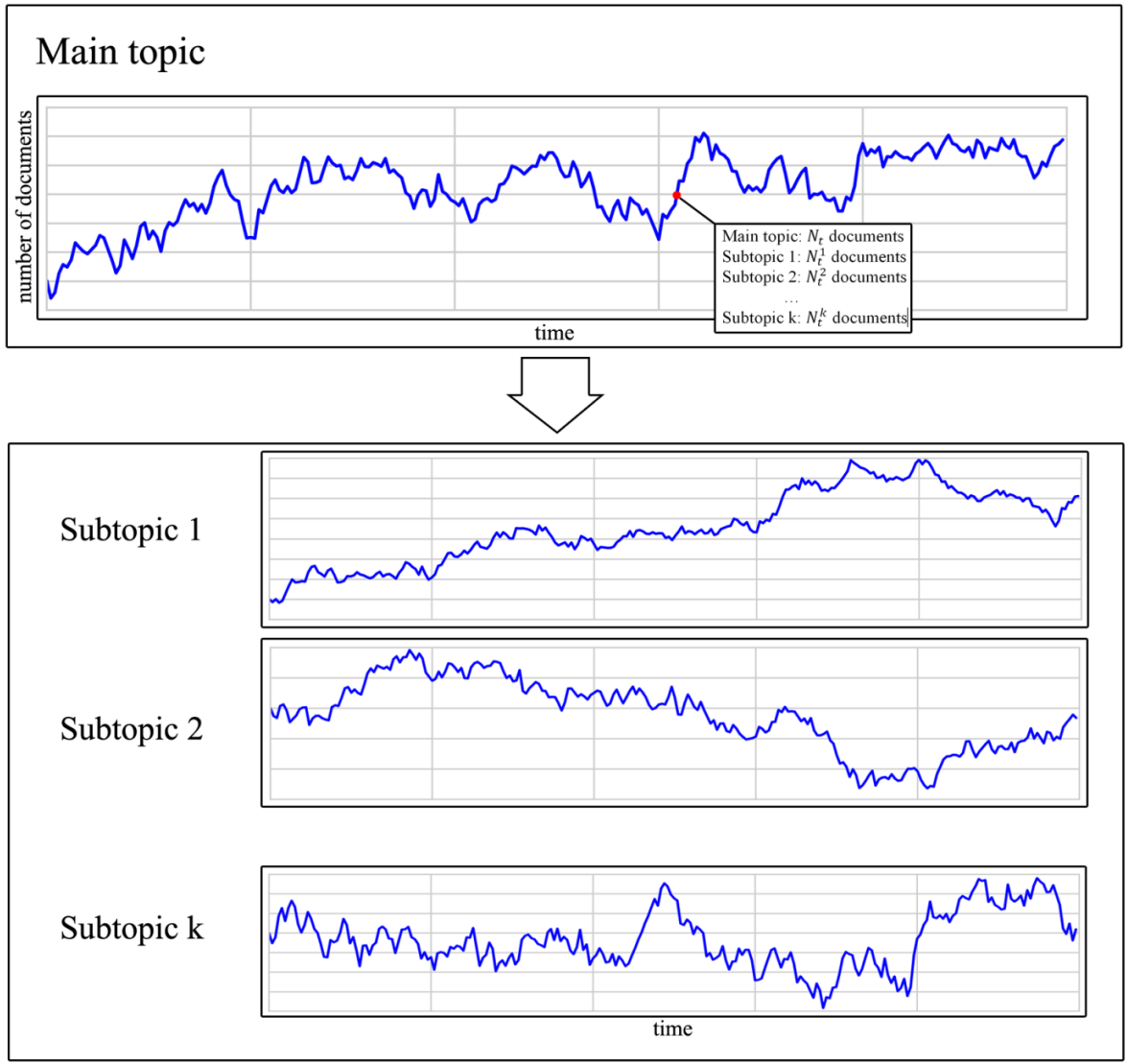

Fig. 1. Schematic illustration of the decomposition

We use following notation and terminology to describe the decomposition. The information stream $S$ is dedicated to the main topic $T$ and has the collection of publications (or documents) $dS$. Each document has a timestamp $t = 1,..., t_{max}$; thus, we can explore the time series $\{x_t\}_{t=1}^{t_{max}}$ (for brevity denote by $x_t$), where $x_t$ is a number of documents with a timestamp $t$. A time unit for information streams is most frequently one day.

First, we consider the simplified case of the decomposition of a stream $S$. We assume the main topic $T$ to have $K$ orthogonal subtopics $T_1,...,T_k$. The orthogonality means that the subtopics are not intersecting. More accurately, each document is corresponding to only one subtopic or does not correspond to any subtopic at all. Therefore, all the documents in $dS$ can be divided into $K+1$ sets:

$$dS = dS_1 \cup dS_2 \cup ... \cup dS_K \cup dS_{other}$$

where $dS_i$ is a document set corresponding to a stream $S_i$. The equality in terms of time series is as follows:

$$x_t = \sum_{i=1}^{K} x_t^{(i)} + x_t^{(other)}.$$

Recall the analogy with the Fourier transform. The Fourier transform provides a coefficient or amplitude for every frequency of oscillation contained in the signal. When applying the decomposition, each subtopic can be matched with a coefficient equaling the number of documents on this subtopic $N_i$ divided by the total number of the documents $N$. This is the way to show the contribution of each subtopic in the whole stream. On the other hand, a contribution of a subtopic may significantly change over time. This leads to the idea to consider the series

$$p_t^{(i)} = \frac{x_t^{(i)}}{x_t},$$

where $\{x_t\}_{t=1}^{t_{max}}$ is the time series corresponding to $S$, and the time series $\{x_t^{(i)}\}_{t=1}^{t_{max}}$ is corresponding to $S_i$.

**2.2 Extracting the main components in the practical case**

To decompose the information stream, it is of great importance to reveal the main components. In practical terms, we got to identify significant subtopics in the text corpora or to construct a domain-specific ontology. Many studies deal with this problem. For example, the technique for producing terminological ontology is presented in [10]. A network of terms is building automatically based on a text corpus. The specific feature of the approach is the usage of a compactified horizontal visibility graphs [11, 12] for terms and connections among them.

Assume that we identify the set of subtopics $T_1,...,T_k$, that are essential for the main topic of the information stream. Now consider the collections of documents $dS_1,...,dS_k$ dedicated to the subtopics. The difference between a practical case and a theoretical one is that the collections of documents have intersections. Therefore, we need to remove duplicates from the union of the sets

$$dS = dS_1 \cup dS_2 \cup ... \cup dS_K \cup dS_{other} \setminus D,$$

where $D$ is the set of duplicates. In terms of time series we have:

$$x_t = \sum_{i=1}^{K} x_t^{(i)} - d_t + x_t^{(other)}.$$

where $d_t$ is the number of duplicated at time $t$.

## 3 Analysis of decomposition results

The result of the decomposition is a set of significant topics and corresponding time series. Further, the following questions arise. What subtopics have the dynamics that is similar to the main time series and what ones are differing significantly? What subtopics may be considered as background? Moreover, in essence, what new useful information can be extracted from the components?

First, it is worth analyzing the principal components separately. The information stream can have unnatural dynamics, which refers to information operations [13, 14]. To perform primary analysis, we use wavelet analysis [15] or such specialized approach as calculating correlation with patterns [14]. These approaches allow us to visualize what parts of the time series resemble the pattern of information operation. In this way, we recognize the segments of the time series that is likely to be influenced by artificial information impact.

To compare the components with the main series, we propose to apply multifractal analysis [16]. Multifractal formalism allows generating multifractal or singularity spectrum based on time series data. The multifractal spectrum is the continuous spectrum of exponents which illustrates scaling properties of the data. Similar spectra indicate similar time series properties. In addition, very narrow spectra indicate weak autocorrelation in series. If it has the spectrum centered around a point, the topic is likely to be noise.

It is not in the scope of this article, but note that multifractal analysis has successfully been applied to diverse applied fields [17-19].

Several methods for estimating the multifractal spectrum have been developed. The most frequently used and mentioned in publications methods are MF-DFA (Multifractal Detrended Fluctuation Analysis)[20] and WTMM (Wavelet Transform Modulus Maxima)[21, 22]. The article [23] is dedicated to the detailed comparison of these methods. The conclusion is that both techniques may be more or less suitable to apply for specific data, but MF-DFA proves to be the universal approach. We also have chosen MF-DFA to use for examples.

We describe the main steps of MF-DFA in brief [20]. To apply the method time series must be aggregated

$$y_t = \sum_{k=1}^{t_{max}} (x_k - \bar{x}), \quad t = 1,...,t_{max}.$$

Determining the aggregated form of a time series is requires because the approach based is on the theory of random walks. Divide the domain into $N_s = \left\lfloor \frac{N}{s} \right\rfloor$ non-overlapping segments of length $s$. If $N$ is not a multiple of $s$, then the same procedure is repeated form the end of the domain ($2N_s$ segments are obtained altogether). Determine the local trend for each segment using a least-square fit. Then calculate the variance

$$F^2(s,\upsilon) = \frac{1}{s}\sum_{i=1}^{s}(y_{(\upsilon-1)s+i} - y_\upsilon(i))^2, \quad \upsilon = 1,...,N_s$$

and the same for $\upsilon = N_s+1,...,2N_s$. Average all the variances to get $q$ th order fluctuation function

$$F_q(s) = \left[\frac{1}{2N_s}\sum_{\upsilon=1}^{2N_s}(F^2(s,\upsilon))^{q/2}\right]^{1/q}, \quad q \in \Re/\{0\}.$$

If the original series has multifractal properties, then the following power-law dependence takes place

$$F_q(s) \sim s^{h(q)},$$

where $h(q)$ is generalized Hurst exponent. After estimating $h(q)$ we can estimate scaling function $\tau(q) = qh(q) - 1$ and, then, multifractal spectrum by applying a Legendre transform

$$\alpha = \tau'(q), \quad f(\alpha) = q\alpha - \tau(q).$$

The estimated multifractal spectra provide an opportunity to compare time series and, in our case, to compare the main information stream and its principal components. By calculating the difference between the main spectrum and subtopics spectra, we range subtopics according to their similarity to the main topic. Therefore, we identify subtopics which evolved similarly to the main topic.

We address one more way to extract valuable information from the principal components. Consider a new stream consisting only of the most significant components. Select those few subtopics (denote the number of them by $K'$) which have the larger number of dedicated documents. Then, consider the new information stream

$$dS = dS_1 \cup dS_2 \cup ... \cup dS_{K'}$$

If the volume of the new stream is sufficient $N' \geq 0.8N$, then analyzing the stream $S'$ instead of $S$ has some advantages. First, all the document in $S'$ deal with at least one subtopic, thus, we reduce the number of insignificant publications. The time series corresponding to $S'$ is, in a sense, 'smoother' compared to $S$. The 'smoothing' is due to deleting insignificant publications and is not related to averaging in common smoothing.

If the volume of the new stream is not sufficient $N' < 0.8N$, then we do not take some significant subtopics in the account. This indicates that determining more subtopics is required. Probably, one should apply computer linguistic methods to extract subtopics from publication texts.

## 4 Example for practical information stream dedicated to Brexit

We present an application of the decomposition to the information stream on the Brexit topic. The dynamics of the number of news publications dedicated to the main topics was gathered using a content-monitoring system. Publishing news is known to have one-week periodicity. To remove periodicity from the time series we divide the values of the series by the total number of documents scanned by the content-monitoring system in a day. Consequently, we use the time series consisted of the relative numbers of publications in a day. Such time series for the Brexit topic is shown in Fig. 2.

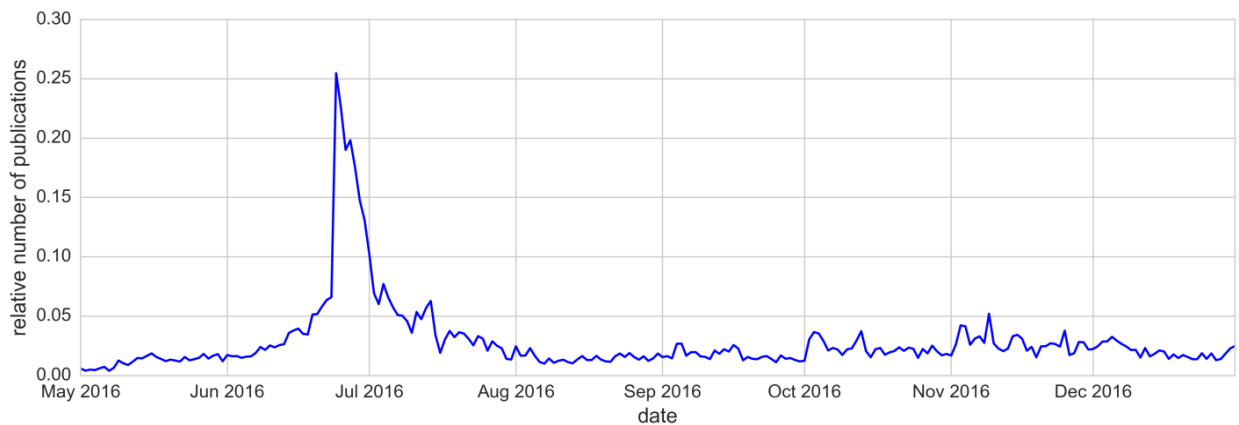

Fig. 2. Relative number of publications in a day dedicated to Brexit from 1.05.16 to 31.12.16

Consider the decomposition for the Brexit stream. In the first place, we choose the topics referring to the aspects influencing votes on the referendum, in accordance with preliminary polls [4, 5]. The following keywords were identifying: trade, immigration, unemployment, taxes. Publications containing these words constitute 41.3% of all documents in the stream.

Having analyzing publications that do not contain such words as immigration, trade, taxes, or unemployment, we identified new keywords characterizing important groups of publications. These words are labour, Tory, referendum, populism, FBI. Each subtopic contributions are summarized in Table 1. Note that 'referendum' is a general word and appear in many publications. On the other hand, 14.6% publications (this number is provided in Table 1) do not refer to any other subtopic besides 'referendum,' and we leave this subtopic from a practical point of view.

Table 1. The percentage of the documents dedicated to subtopics

| Topic | Percent of publications |
|---|---|
| Trade | 28.4% |
| Immigration | 20.6% |
| Labour | 15.4% |
| Referendum | 14.6% |
| Tory | 8.8% |
| Unemployment | 3.4% |
| Taxes | 3.3% |
| Populism | 2.0% |
| FBI | 1.3% |

Besides the contribution summarized over the whole time, it is essential to consider how the contribution of a subtopic varies. Shown in Fig. 3 is the contributions of trade, immigration, and FBI depending on time. One can notice that issues connected with 'trade' were discussing all the year round, but the FBI topic arose only in November.

To compare the principal components and the main series we have estimated the multifractal spectra discussed in Section 4. Some of the spectra are shown in Fig. 4. Worth emphasizing that time series dedicated to FBI and Populism have too many zeros to estimate the multifractal spectra because both topics arisen at the end of 2016. The standard deviations between the spectra have been calculated for other topics. The results are presented in Table 2. The conclusion is that the Tory ad Trade topics are the most similar to the Brexit topic in terms of comparing multifractal spectra.

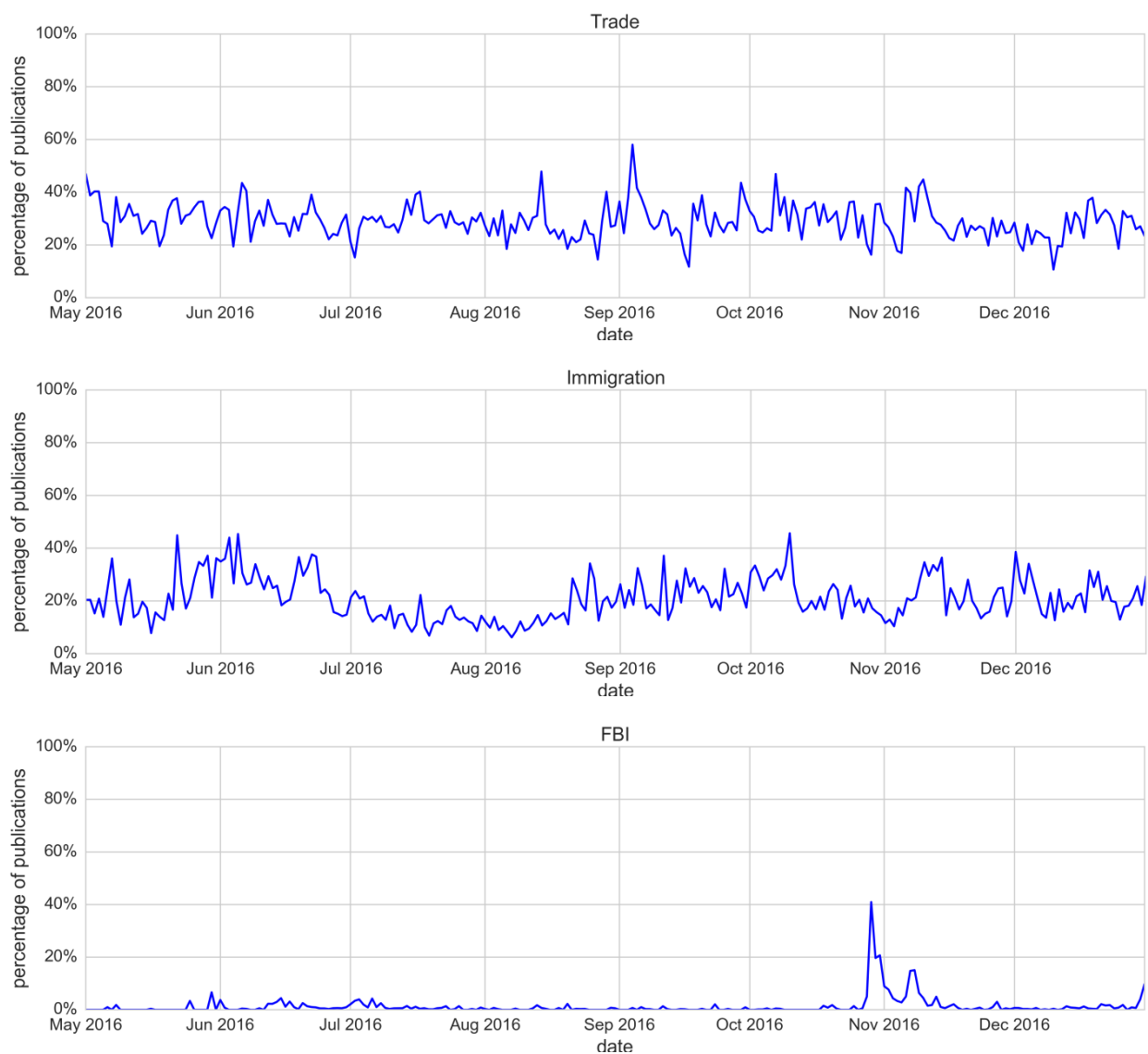

Fig. 3. The number of publications on trade, immigration, FBI divided by the total number of publications on Brexit in a day

**5 Discussion and conclusion**

Finally, we address a comparison of the principal components obtained by applying the decomposition to the Brexit topic and topics related to Brexit according to Google Trends. We have gathered Google Trends data about the topics related to Brexit in 2016 and show it in Table 3 and 4.

The topic, which is shared by the decomposition results and the related topics from Google Trends, is 'referendum.' Apparently, it is the most general topic related to Brexit. Other obtained topics differ significantly. Google Trends identified several names belonging to prominent British politicians. Note that Boris Johnson, David Cameron, and Michael Gove are members of the Conservative Party that appeared in the decomposition. Similarly, the related topics include economic terms such as Pound sterling, Euro, Economy, which are referred to the Trade topic from the decomposition. Therefore, some identified topics are presented by different keywords but are similar in both approaches.

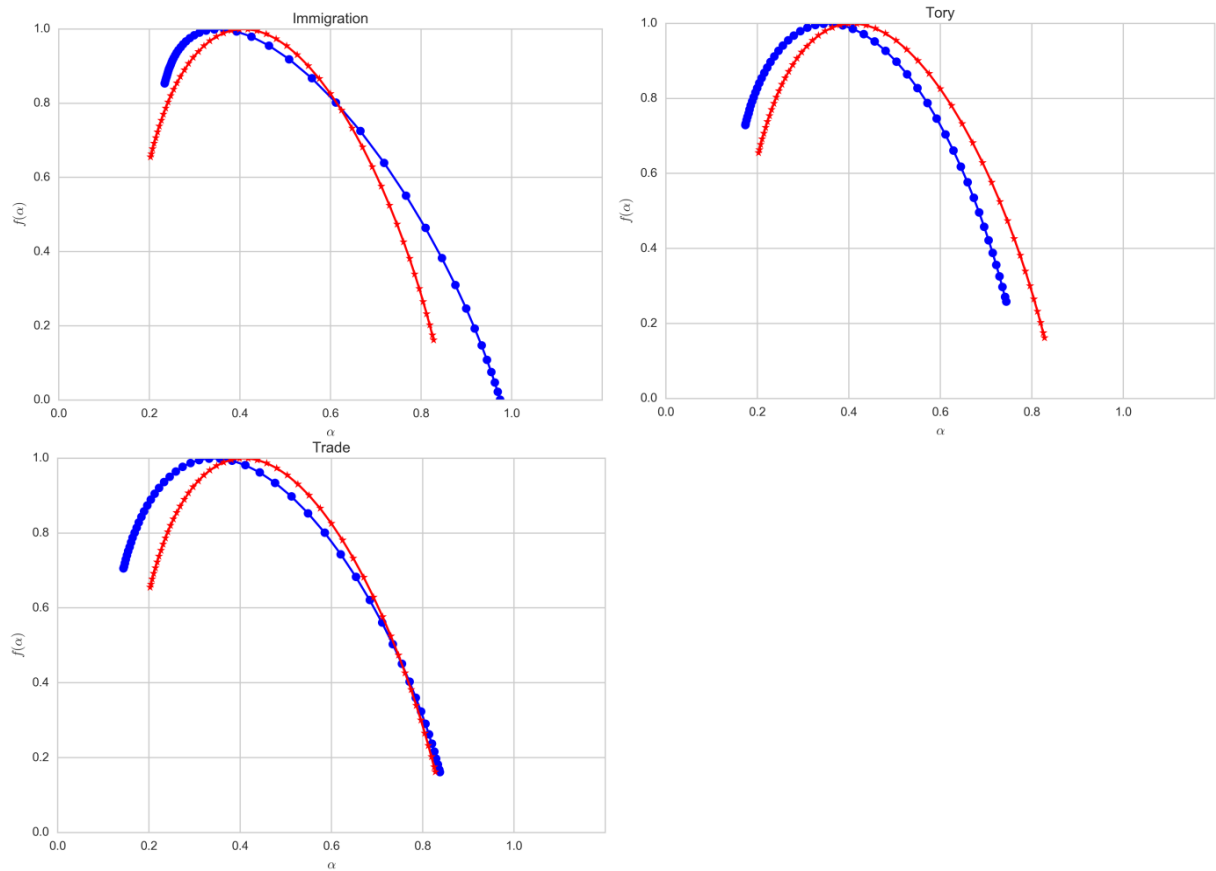

Fig. 4. The multifractal spectra for the Brexit time series (stars) and the Immigration, Tory и Trade series (circles)

Table. 2. The standard deviations between the main spectrum and the principal components spectra

| Topic | Standard deviation |
|---|---|
| Tory | 0.047 |
| Trade | 0.056 |
| Referendum | 0.083 |
| Immigration | 0.106 |
| Taxes | 0.113 |
| Labour | 0.133 |
| Unemployment | 0.172 |
| Populism | - |
| FBI | - |

Table 3. The topics related to Brexit from Google Trends

| The most popular topics | | Topics with the biggest increase in search frequency | |
|---|---|---|---|
| Topic | Relative scale | Topic | Increase |
| Opinion poll | 100 | Exit poll | >1000% |
| United Kingdom European Union membership referendum, 2016 | 67 | Pound sterling | >1000% |
| Referendum | 44 | England | >1000% |
| European Union | 13 | Voter turnout | >1000% |
| Boris Johnson | 8 | Scotland | >1000% |
| David Cameron | 8 | Petition | >1000% |
| Europe | 7 | Michael Gove | >1000% |
| Nigel Farage | 7 | Resignation | >1000% |
| Exit poll | 6 | United Kingdom European Union membership referendum, 2016 | 950% |
| Survey methodology | 6 | Opinion poll | 750% |
| Pound sterling | 5 | Boris Johnson | 700% |
| Opinion | 5 | Survey methodology | 450% |
| England | 4 | Referendum | 250% |
| Voter turnout | 3 | Nigel Farage | 200% |
| Scotland | 3 | Opinion | 140% |
| Petition | 3 | Europe | 120% |
| Michael Gove | 2 | David Cameron | 50% |
| Resignation | 2 | | |

By decomposing, we also detect topics which were popular during the short period of time in 2016 such as Populism and FBI. Naturally, Google Trends did not identify these topics as the most popular during the year. On the other side, the principal components do not include Survey methodology, Resignation, and Brexit: The Movie.

To summarize, we have proposed the approach to decomposing the information stream into the principal components connected to the most significant topics of the documents. This approach allows reducing the topical issue of automated processing information streams. The decomposition can be applied to detecting primordial and secondary topics, as well as finding unimportant documents. Analyzing each principal component separately provides additional information to recognize critical moments of external impact. Therefore, the decomposition can be applied to the computational and analytical processing of information streams. Further, the obtained results can be used in a decision support system.

Table 4. The topics related to Brexit in the News category from Google Trends

| The most popular topics | | Topics with the biggest increase in search frequency | |
|---|---|---|---|
| Topic | Relative scale | Topic | Relative scale |
| United Kingdom | 100 | Result | >1000% |
| Opinion poll | 72 | Theresa May | >1000% |
| Voting | 60 | Petition | >1000% |
| European Union | 58 | Racism | >1000% |
| United Kingdom European Union membership referendum, 2016 | 55 | Brexit: The Movie | >1000% |
| Referendum | 45 | Prime minister | >1000% |
| Result | 41 | Member of Parliament | >1000% |
| Pound sterling | 31 | Voter turnout | >1000% |
| Europe | 31 | Pound sterling | 850% |
| Euro | 22 | Voting | 350% |
| Odds | 19 | Opinion poll | 300% |
| England | 18 | Scotland | 250% |
| Scotland | 12 | Odds | 250% |
| BBC News | 10 | United Kingdom European Union membership referendum, 2016 | 200% |
| Economy | 10 | BBC News | 150% |
| The Guardian | 9 | Economy | 80% |
| Ireland | 8 | Referendum | 70% |
| Theresa May | 7 | England | 60% |
| Petition | 5 | United Kingdom | 50% |
| Racism | 3 | Euro | 50% |
| Brexit: The Movie | 3 | | |